\documentclass{article}
\usepackage{graphicx}
\usepackage{amsmath}
\usepackage{amssymb}
\usepackage{rotating}
\usepackage{epsfig}
\usepackage{verbatim} 
\usepackage{a4}
\input amssym.def
\input amssym.tex
\def\ben{\begin{equation}}
\def\een{\end{equation}}

\def\bea{\begin{eqnarray}}
\def\eea{\end{eqnarray}}

\def \p{{\bf p}}

\def \p{\partial}

\title{A Note on the Instability of Lorentzian Taub-NUT-Space}
\date{February 13, 2006}
\author{Gustav Holzegel\thanks{G.Holzegel@damtp.cam.ac.uk} \\
University of Cambridge, DAMTP, Wilberforce Road, \\
Cambridge CB3 0WA, United Kingdom}

\begin{document}

\maketitle

\begin{abstract}
I show that there are no $SU(2)$-invariant (time-dependent) 
tensorial perturbations of Lorentzian Taub-NUT space. It follows that
the spacetime is unstable at the linear
level against generic perturbations. I speculate that this 
fact is responsible for so far unsuccessful 
attempts to define a sensible thermodynamics for NUT-charged 
spacetimes.
\end{abstract}

\section{Introduction}
There has been some interest recently in spacetimes admitting
NUT-charge and the thermodynamics associated to them, 
see e.g. \cite{Mann, Mann1, Mann2} and references therein. 
Inspired by the AdS-CFT-conjecture the entropy for many
 such spacetimes has been computed by counter-term methods
but a satisfactory first law of thermodynamics has not yet been
stated. 
Given the confusion about the thermodynamics of
NUT-spacetimes, and in fact about NUT-charge in general, 
it is perhaps illuminating to perform a Lorentzian linear stability
analysis for these spacetimes. In this paper the
simplest case, Lorentzian-Taub-NUT without a cosmological constant, is
addressed, finding instability of the spacetime at the linear level. 
\\

I revisited a paper \cite{Young}, which provides the methods to 
analyse linear stability of spacetimes admitting an $SU(2)$ 
invariance by decomposition of the perturbation
into tensor-harmonics using Wigner-functions. 
In \cite{Young} the signature used was
$(+,+,+,+)$ and the stability of gravitational instantons was 
analyzed. These results have recently been generalized to 
non-vanishing cosmological constant by Warnick \cite{Claude}. 
However, the same techniques can be applied if the signature is 
Lorentzian. Looking only at the lowest tensor-harmonic, this
leads to the result proven in this paper, namely 
that there exist no $SU(2)$-invariant 
tensor-perturbations of Taub-NUT space. This means that possible 
perturbations behave either badly at infinity or at the horizon 
in a sense that will be made precise below. 
It is important to note that the perturbations 
under consideration are \emph{time-dependent}, with a periodicity 
dictated by the $SU(2)$ symmetry of the metric. Having found this kind
of instability in the lowest mode, it follows that the spacetime 
is unstable at the linear level against generic perturbations. 
\\

It is interesting to relate the result to the full non-linear
problem in the $SU(2)$-symmetry class, which corresponds to finding 
solutions to the Einstein equations for the following metric:
\begin{equation} \label{prob}
ds^2 = f(r)dr^2 + A(r)\sigma_1^2 + B(r)\sigma_2^2  - C(r)\sigma_3^2 \, .
\end{equation}
with the $SU(2)$ left-invariant one-forms $\sigma_i$ defined in equation
(\ref{liforms}) and for positive functions $f,A,B$ and $C$.
The metric (\ref{prob}) is related to the famous,
thoroughly studied, Bianchi-IX cosmologies for which the $SU(2)$ 
acts on \emph{spatial} hypersurfaces. The metric
(\ref{prob}) has also been studied intensively with Euclidean signature 
in the search for gravitational instantons. In both cases
 the Einstein-equations can be solved exactly only under 
restrictive conditions like $A(r)=B(r)$. In the case (\ref{prob})
considered here there will be important sign changes 
compared to either Bianchi-IX or Euclidean signature. 
Hence an interesting problem, worth to be 
studied in its own right, is to resolve whether the assumption 
of $SU(2)$ invariance in the form of the metric ansatz (\ref{prob}) 
already forces the spacetime to be stationary, as might be suggested 
by the result of this paper. This would correspond to a 
Birkhoff-type theorem. Note that the standard Birkhoff's theorem is a local
statement about any piece of spacetime, with the symmetry group $SO(3)$ 
acting on a two-sphere. Here, on the other hand, the
$SU(2)$ acts on a three-sphere and its action implies in particular the 
existence of closed timelike curves. Therefore it is the global
nature of the spacetime, which might make a ``no-dynamics'' 
result possible.
\\

For Lorentzian Taub-NUT there is in fact a rigorous theorem for the non-linear
problem. The authors of \cite{Moncrief} (see also \cite{Gibbons}) 
prove that any spacetime admitting a compactly generated Cauchy 
horizon with closed generators, necessarily 
has a Killing-symmetry, which changes from being spacelike to timelike 
when crossing the Cauchy-horizon. In other words, situations as they 
occur in Taub-NUT are essentially an artefact of symmetry. In
particular, there is 
no dynamics of the Cauchy horizons (i.e. $dA=0$), and hence no
standard first law of thermodynamics to which physical meaning 
could be associated. 
The arguments of \cite{Moncrief} require the 
full non-linear Einstein equations. While they make 
no explicit connection with stability, they certainly suggest that 
the Taub-NUT metric is dynamically uninteresting. 
The result of the present paper 
demonstrates that the pathological behaviour of 
Lorentzian Taub-NUT-space already shows up at the linear level 
by considering the stability of the metric under 
perturbations. 
\\

Let me finally comment on the issue of thermodynamics of 
NUT-charged spacetimes. In most recent attempts \cite{Mann1} 
some first law was imposed a priori from Euclidean quantum gravity
arguments. However, this assumption enforced an additional, 
physically uninterpreted, relationship between the NUT-charge 
and the horizon-radius, leaving the physical significance 
of the first law somewhat dubious. On the other hand, one could 
stick to the view that a truely satisfying first-law has to 
determine the intrinsic and extrinsic variables, and state 
explicitly which physical quantities can be varied in what
situation. An attempt in that direction is outlined in the appendix. 
There it turns out that a NUT-potential can be derived 
and that in general a Smarr-formula holds for Kerr-Taub-NUT spacetime. 
However, a first-law can not be deduced from the Smarr-formula in
the usual fashion. It is perhaps instructive to see how this attempt 
fails: Similarly to \cite{Mann1} there seems 
to be a missing degree of freedom. I speculate that all these 
difficulties are rooted in the Lorentzian instability 
of the spacetime.

\section{The geometry of Taub-NUT spacetime}
The four-dimensional Taub-NUT spacetime is given by the metric
\begin{equation} \label{nutmetric}
g^{TN} = -A(r) (dt + 2n \cos \theta d \phi) ^2+ 
\frac{dr^2}{A(r)} +B(r)(d \theta ^2 + \sin ^2 \theta d \phi ^2 )\, ,
\end{equation}
where
\begin{align}
A(r) &= \frac{r^2-2mr-n^2}{r^2+n^2} \, , \\
B(r) &= r^2+n^2 \, .
\end{align}
The metric has singularities where $A(r)=0$ and where $\theta=0,
\pi$. The former correspond to horizons at $r_{\pm}= m \pm
\sqrt{m^2+n^2}$.
The singularities at $\theta=0,\pi$ are not the usual coordinate 
degeneracies of the two sphere. In fact if we allow $-\infty<t<\infty$ 
the metric (\ref{nutmetric}) will have a line-singularity 
along the $z$-axis, whose existence needs to be interpreted 
physically in one or the other way (see \cite{Bonnor} and \cite{Manko} for 
two different suggestions). On the other hand, if the coordinate 
$t$ is assumed to be periodic with period $8 \pi n$, 
the singularities in $\theta$ can be interpreted as the usual
Euler-\emph{coordinate} singularities of the three-sphere, as was
first observed by Misner \cite{Misner}. 
\\
In this case spacetime is covered by two coordinate patches
The first patch covers a region around the north pole ($\theta=\pi$)
namely the region $\theta \geq \frac{\pi}{2}$. 
This is achieved by the coordinate transformation 
$t \mapsto t^{(+)} = t + 2n \phi$ which results in
\begin{equation}
ds_{(+)}^2 = -A(r) \left[dt^{(+)} - 2n\left(1-\cos \theta\right)
  d\phi \right]^2 + \frac{dr^2}{A(r)} + 
B(r)(d \theta ^2 + \sin ^2 \theta d \phi ^2 ) \, ,
\end{equation}
a metric regular around the north pole.
For the second patch around the south pole, $\theta=0$, we have to use the
coordinate transformation $t \mapsto t^{(-)} = t - 2n \phi$, which
leaves us with 
\begin{equation}
ds_{(-)}^2 = -A(r) \left[dt^{(-)} - 2n\left(1+\cos \theta\right)
  d\phi \right]^2 + \frac{dr^2}{A(r)} + 
B(r)(d \theta ^2 + \sin ^2 \theta d \phi ^2 ) \, ,
\end{equation}
a metric regular in the region $\theta \leq \frac{\pi}{2}$.
In the overlap, which is the equator, $\theta=\frac{\pi}{2}$, we have
\begin{equation}
t^{(+)} = t^{(-)} + 4n \phi \, ,
\end{equation}
and therefore consistency requires the coordinates $t^{(+)}$ 
and $t^{(-)}$ to be periodic with a period of $8\pi n$. 
\\
In this paper we will follow \cite{Misner} and assume periodic
$t$-coordinate to avoid the line-singularity. Our spacetime 
then has topology $\mathbb{R} \times S^3$, it is free 
of singularities but there are closed timelike curves for $A(r)>0$.

\section{Stability}
Let us assume a perturbation of the Taub-NUT metric (\ref{nutmetric}):
\begin{equation}
g_{ab} = g_{ab}^{TN} + h_{ab} \, .
\end{equation}
Fixing the gauge we require the perturbation $h_{ab}$ 
to be transverse with respect to the Taub-NUT derivative operator
\begin{equation} \label{transverse}
\nabla^{a} h_{ab} = 0 \, ,
\end{equation}
and traceless
\begin{equation} \label{traceless}
g^{TN}_{ab} h^{ab} = 0 \,  .
\end{equation}
Furthermore, it should -- at least to first order -- 
satisfy the vacuum Einstein-equations
\begin{equation}
R_{ab}(g)=0 \, .
\end{equation}
This implies the following equation for the perturbation $h_{ab}$:
\begin{equation} \label{stab}
\bigtriangleup_{L}h_{ab} =-\nabla^e \nabla_e h_{ab} - 2R_{acbd}h^{cd}
=  0 \, ,
\end{equation}
where we have defined the the Lichnerowicz-operator $\bigtriangleup_L$.
All derivative operators are taken with respect 
to the background Taub-NUT metric. 
As a last requirement, the perturbation should be finite in the sense
that the invariant expression
\begin{equation} 
tr \left(g\right)=g_{ab}g^{ab}=\left(g_{ab}^{TN}+h_{ab}\right)\left(g^{TN \ ab}+h^{ab}\right)=4
+ h_{ab}h^{ab} 
\end{equation}
is bounded everywhere between the horizon and infinity. That is 
\begin{equation} \label{bound}
h_{ab}h^{ab} <\infty \textrm{\ \ \ everywhere between the horizon and infinity.}
\end{equation}
Of course, stronger notions of finiteness and smallness could be
introduced but the condition (\ref{bound}) will suffice for our
purposes because possible perturbations will already violate this 
elementary condition.

Summarizing, we want perturbations $h_{ab}$ satisfying
 (\ref{transverse}), (\ref{traceless}), and (\ref{stab}) and being
finite in the sense of (\ref{bound}).
In general such solutions are hard to find. The key is to 
exploit the $SU(2)$-invariance of the background
by expanding the perturbations in terms of Wigner-functions
(``modes''). If we then restrict ourselves to $SU(2)$-invariant
perturbations (that is, zero modes, $J=K=M=0$),
we can prove the following result
\\
\\
\emph{There are no finite, transverse, traceless 
$SU(2)$-invariant perturbations of the \\
Lorentzian Taub-NUT metric.}
\\
\\
This result indicates that the Taub-NUT-metric is unstable. 
As mentioned in the introduction this result regarding linear
 stability is related to the non-linear ``no-dynamics''-result 
proved in \cite{Moncrief}.  \\
We will make further comments about the type of the perturbation after 
presenting its precise form in the next section.

\section{Decomposition of perturbations}
The Taub-NUT metric has an $SU(2)$ invariance, which can be made more
manifest by introducing the left invariant one-forms of $SU(2)$
(coordinatized by Euler-coordinates $\theta, \phi, t$)
\begin{align} \label{liforms}
\begin{split}
\sigma_1 &= -\sin t d\theta + \sin \theta \cos t d\phi \, ,\\
\sigma_2 &= \cos t d\theta + \sin \theta \sin t d\phi \, , \\
\sigma_3 &= dt + \cos \theta d\phi \, ,
\end{split}
\end{align}
and writing the metric (\ref{nutmetric}) as
\begin{equation}
ds^2= \frac{1}{A(r)} dr^2 - 4n^2 A(r) \sigma_3^2 + B(r)
\left(\sigma_1^2 + \sigma_2^2 \right) \, .
\end{equation}
For future reference we also define
\begin{align} \label{gamma}
\begin{split}
\gamma_{00} &= \gamma_{rr} = \frac{1}{A(r)} \, , \\
\gamma_{11} &= \gamma_{22} = B(r) \, , \\
\gamma_{33} &= - 4n^2 A(r) \, .
\end{split}
\end{align}
As mentioned in the last section, the key is an expansion of the
perturbations in terms of $SU(2)$-Wigner-functions
$\mathcal{D}_{KM}^{\phantom{KM}J}$, which are analogues of the
spherical harmonics on the two-sphere. The general perturbation 
can be decomposed as 
\begin{equation} \label{decomp}
h = \sum_{J,M} \sum_{K=-J}^{K=J} h^{J,M,K}_{ab}(r)
\mathcal{D}_{KM}^{\phantom{KM}J}\left(t,\theta,\phi \right)
 \sigma^a \otimes \sigma^b \, .
\end{equation}
The tedious work of rephrasing the equations (\ref{transverse}),
(\ref{traceless}), and (\ref{stab}) in terms of modes is done 
in \cite{Young}, where the \emph{Euclidean} Taub-NUT metric is
studied. In particular, the zero-mode equations for 
$h^{0,0,0}_{ab}$ together with their transverse-traceless gauge conditions 
are derived. Taking care of the signs all these equations can be restated 
for the Lorentzian case. In both cases second order ordinary
differential equations can be decoupled for the components 
$h_{00}, h_{33}, h_{03}, X=h_{11}+h_{22}, Y=h_{11}-h_{22}, h_{12}, 
h_{13}, h_{02}, h_{01}, h_{23}$ of the perturbation $h_{ab}$. 
Here and in the following we suppress the $(0,0,0)$-mode-indices.
The strategy followed in this paper will be to analyze the equation 
for each component carefully and to prove that the solutions they
 admit will behave badly at the horizon or at infinity. 
To be more specific we will show that any 
solution would violate the boundedness condition (\ref{bound}).  
\\
Note that the zero mode under consideration will in general be
time-dependent. This is because the the zero mode of (\ref{decomp})
 is of the form
\begin{equation}
h = h^{0,0,0}_{ab}(r) \sigma^a \otimes \sigma^b \, ,
\end{equation}
using that $\mathcal{D}_{00}^{\phantom{00}0}=const$. 
Since the left invariant one-forms (\ref{liforms}) 
explicitly depend on time in a periodic fashion, 
the modes under consideration are periodic in time. It is this fact
that renders the result non-trivial.
\\
It is also worth pointing out that our non-existence result is 
different in style from the usual linear perturbation-theory arguments 
for spacetimes without closed timelike curves but an ordinary timelike
Killing vector (see e.g. \cite{Sean}). There it is usually shown that 
regular perturbations \emph{can} be defined on a spacelike
hypersurface but that they will blow up in time. Here we show that 
in fact no $SU(2)$-invariant perturbation can be defined initially 
that behaves well at the horizon and at infinity.

\section{Differential equations}
In the following, we will constantly refer to equations derived in
\cite{Young}. The differential equations for the metric perturbation
components $h_{ab}$ stated in this section can be derived from
the Wigner-decomposition and are found explicitly (with Euclidean
signature) in the appendix of that paper.
\subsection{$h_{21}$}
The $h_{21}$ equation is ((B1) of \cite{Young})
\begin{equation}
a(r)h_{21}^{\prime \prime}(r) + b(r)h_{21}^\prime(r) + c(r)h_{21}(r)=
0 \, ,
\end{equation}
where,
\begin{align}
a(r)&=A(r) \, , \\
b(r)&=A^\prime - \frac{AB^\prime}{B} \, , \\
c(r)&=\frac{1}{n^2A}-\frac{A^\prime
  B^\prime}{B}+A\left(\frac{B^{\prime}}{B}\right)^2-A\left(\frac{B^{\prime \prime}}{B}\right) + \frac{2}{B} \, .
\end{align} 
Near infinity the equation reads
\begin{equation}
h_{21}^{\prime \prime} + \left(-\frac{2}{r} +
\mathcal{O}\left(\frac{1}{r^2}\right) \right) h^\prime_{21} +
\left(\frac{1}{n^2} + \mathcal{O}\left(\frac{1}{r}\right)\right)
h_{21} = 0 \, ,
\end{equation}
singnalizing an irregular singularity near infinity. 
Using techniques from ordinary differential equations we can 
compute an asymptotic expansion
near infinity. The two linearly independent asymptotic solutions are
\begin{align} \label{h21sol}
h_{21} &= -\sqrt{\frac{2n}{\pi}}\left(r\cos\left(\frac{r}{n}\right) -
n \sin \left(\frac{r}{n} \right) \right)
\left(1+\mathcal{O}\left(\frac{1}{r}\right)\right) \, ,
\\
h_{21} &= -\sqrt{\frac{2n}{\pi}}\left(n\cos\left(\frac{r}{n}\right) +
r \sin \left(\frac{r}{n} \right)
\right)\left(1+\mathcal{O}\left(\frac{1}{r}\right)\right) \, .
\end{align}
None of these solutions decays at infinity, which is a necessary
requirement for the perturbation. Hence $h_{21}=0$. Note that it is
the Lorentzian signature which is responsible for the oscillatory
behaviour. For the Euclidean version one would have to replace
$\frac{1}{n^2}$ by $-\frac{1}{n^2}$ and exponentially decaying
solutions will exist. (Whether they are well behaved at the horizon
remains to be checked.) This behaviour, that is the asymptotics of the
perturbation being governed by the signature of the background, will
appear in almost all of the following differential equations 
we are about to consider.

\subsection{$h_{20}$ and $h_{10}$}
The components $h_{20}$ and $h_{10}$ satisfy the same differential
equation. We spell out the formulae for $h_{20}$ here but all
conclusions are of course valid for $h_{10}$ as well. \\
(Inserting equations (B6) into (B4) of \cite{Young}, respectively (B9)
into (B7) yields)
\begin{equation}
a(r) h^{\prime \prime}_{20}(r) +
\left(b(r)-\frac{Ad(r)}{\frac{1}{2B}+\frac{1}{8n^2A}}\right)h^\prime_{20}(r)+
\left(c(r)-d(r)\frac{A^\prime+\frac{AB^\prime}{B}}{\frac{1}{2B}+\frac{1}{8n^2A}}\right)h_{20}(r)
  = 0 \, ,
\end{equation}
with,
\begin{align}
a(r) &= -A \, , \\
b(r) &= -2 A^\prime \, , \\
c(r) &=
\frac{3}{2}A\left(\frac{B^\prime}{B}\right)^2-\frac{1}{4n^2A}-\frac{2n^2A}{B^2}-\frac{1}{2}A
\left(\frac{B^{\prime \prime}}{B}\right)-\frac{1}{B}-\frac{A^{\prime
    \prime}}{2} \, , \\
d(r) &=
\frac{B^\prime}{B^2}-\frac{A^\prime}{2AB}+\frac{A^\prime}{8n^2A^2} \, .
\end{align}
Near infinity the differential equation looks like
\begin{equation}
h^{\prime \prime}_{20}(r) + \left(\frac{6m}{r^2} + \frac{8m^2+28n^2}{r^3}
+ \mathcal{O}\left(\frac{1}{r^4}\right)\right) h_{20}^\prime(r) +
\left(\frac{1}{4n^2}+\mathcal{O}\left(\frac{1}{r}\right)\right)
h_{20}(r) = 0 \, ,
\end{equation}
and we can construct the asymptotic solutions by applying theorems about
irregular singular points at infinity. The two solutions are
\begin{equation}
h_{20}(r) = \exp\left(\frac{i}{2n}r\right) \left[1
+\mathcal{O}\left(\frac{1}{r}\right) \right] \, ,
\end{equation}  
and its complex conjugate. Note that the series in the square-brackets
may have complex coefficients. We can easily construct two real
solutions and it is immediate that they will be oscillatory near
infinity. Hence $h_{20}=h_{10}=0$.

\subsection{$Y=h_{11}-h_{22}$}
An equation for the difference $Y=h_{11}-h_{22}$ can be decoupled from
equations (B10) and (B11) of \cite{Young}. It reads
\begin{equation}
a(r)Y^{\prime \prime}(r) + b(r)Y^\prime(r) +
\left(d(r)-e(r)\right)Y(r)=0 \, ,
\end{equation}
where
\begin{align}
a(r) &= A(r) \, , \\
b(r) &= A^\prime - A\frac{B^\prime}{B} \, , \\
d(r) &= \frac{2}{B} - \frac{AB^{\prime \prime}}{B} - \frac{A^\prime
  B^\prime}{B} +
\frac{1}{2}A\left(\frac{B^\prime}{B}\right)^2+\frac{4n^2A}{B^2} +
\frac{1}{n^2A} \, , \\
e(r) &= -\frac{1}{2n^2A} + \frac{4n^2A}{B^2} -
\frac{1}{2}A\left(\frac{B^\prime}{B}\right)^2 \, .
\end{align}
Near infinity the differential equation reads
\begin{equation}
Y^{\prime \prime}(r) + \left(-\frac{2}{r} +
\mathcal{O}\left(\frac{1}{r^2}\right) \right) Y^\prime(r) +
\left(\frac{3}{2n^2} +
\mathcal{O}\left(\frac{1}{r}\right)\right) Y(r) = 0 \, .
\end{equation}
The asymptotic structure is very similar to that of the $h_{21}$
equation considered above. In fact the arguments
used in that subsection can be brought forward to show that solutions
of this equations have to oscillate (and blow up) at infinity. This
shows that $Y=0$ and therefore $h_{11}=h_{22}$.

\subsection{$h_{30}$}
The solution for $h_{30}$ can be obtained explicitly 
from the contraint equation ((B2) of \cite{Young})
\begin{equation}
h_{30}^\prime = -\left(\frac{A^\prime}{A}+\frac{B^\prime}{B}\right)
h_{30} \, .
\end{equation}
It is
\begin{equation} \label{h30}
h_{30}(r) = \frac{1}{A(r)B(r)} = \frac{1}{r^2-2mr-n^2} \, .
\end{equation}

\subsection{$h_{00}$}
The $h_{00}$ equation that can be decoupled from (B14), (B15), (B12)
of \cite{Young} is the most complicated one. It reads
\begin{equation}
\label{diffeq}
a(r) h''_{00} + b(r) h'_{00} + c(r) h_{00} = 0 \, ,
\end{equation}
with the coefficients given by
\begin{eqnarray*}
a(r) &=& A \, , \\
b(r) &=& 3 A' + \frac{AB'}{B} + \frac{B(A'^2 - 2A
  A'')}{B A' - A B'} - \frac{A}{B} \left( \frac{2AB'^2 -
  BA'B' - 2ABB''}{B A' - A B'} \right) \, , \\
c(r) &=&  \frac{A'^2}{2A} + \frac{A'B'}{B} -
\frac{AB'^2}{B^2} + A'' + \frac{3}{2A} \left( \frac{BA'
  + A B' }{BA' - AB'}\right) \\ && \hspace{.5cm} + \frac{B A'B' +
  2BAB'' - 2AB'^2}{B^2} \left( \frac{2BA'
  + A B' }{BA' - AB'}\right) \, .
\end{eqnarray*}
Inserting the functions $A$ and $B$ the equation becomes
\begin{equation}
h^{\prime \prime}_{00}(r) + p(r) h^\prime_{00}(r) + q(r) h_{00}(r) = 0
\, ,
\end{equation}
where now
\begin{align}
p(r) &= \frac{8\left( m - r \right) }{n^2 + \left( 2m - r \right)
  r} + \frac{2r}{n^2 + r^2} + 
  \frac{6\left( n^2 + \left( 2m - r \right) r \right)
  }{-3n^2r + r^3 + m\left( n^2 - 3r^2 \right) } \, , \\
q(r) &= 2\Bigg( \frac{4\left( m^2 + n^2 \right) }{{\left( n^2 + 
\left( 2m - r \right) r \right) }^2} + \frac{2n^2}{{\left( n^2 +
  r^2 \right) }^2} + \frac{1}{n^2 + r^2} + \\
 & \ \ \ \ \ \ \ \ 
\frac{8}{-n^2 - 2mr + r^2} + \frac{9\left( m - r \right)
  }{-3n^2r + r^3 + m\left( n^2 - 3r^2 \right) } \Bigg) \, .
\end{align}
There are (regular) singular points at the horizon, $r_{hoz}=m +
\sqrt{m^2+n^2}$, and at infinity. There is also a regular singular point
at
\begin{equation}
r=r_{int} = m +
2\cos\left(\frac{\arctan\left(\frac{n}{m}\right)}{3}\right) 
\sqrt{m^2+n^2} \, .
\end{equation}
However, a Frobenius expansion around $r_{int}$ 
shows that the solution has always finite value of the 
first and second derivative at that point. \\
The regular singular point at infinity can also be analyzed by a Frobenius
series. Near infinity 
\begin{align}
p(r) &= \frac{4}{r} + 2m \frac{1}{r^2} +
\mathcal{O}\left(\frac{1}{r^3}\right) \, , \\
q(r) &= -\frac{4m}{r^3} + \mathcal{O}\left(\frac{1}{r^4}\right) \, ,
\end{align}
therefore the solution is asymptotically either  $h_{00} \propto \frac{1}{r^3}$ or
proportional to a constant $(\neq0)$ near infinity. Of course, 
only the first case will lead to a sensible perturbation.
In fact, the solution $h_{00}$ that behaves like $\frac{1}{r^3}$ near infinity
can be constructed explicitly and is unique by the theory of ODEs. It reads
\begin{equation} \label{h00}
h_{00} =
\frac{2r^3-m\left(n^2+3r^2\right)}{\left(r^2+n^2\right)
  \left(r^2-2mr-n^2\right)^2} \, .
\end{equation}

\subsection{The other components}
So far we found that for any finite, transverse, traceless perturbation,
$h_{21}=h_{20}=h_{10}=0$ must hold. Using the equations given in \cite{Young}
one concludes that 
\begin{itemize}
\item $h_{20}=0$ implies $h_{31}=0$ (by (B6) of \cite{Young})
\item $h_{10}=0$ implies $h_{32}=0$ (by (B9) of \cite{Young})
\end{itemize}
Thus there are only five possibly non-zero components left, 
of which we already know $h_{00}$
and $h_{30}$ explicitly. The remanining components could be calculated
explicitly from what we already know but in fact there is no need to
do this. Writing out the boundedness condition (\ref{bound}) yields
\begin{equation} \label{finit}
h_{ab}h^{ab} = h_{00}h^{00} + 2h_{30}h^{30} + h_{11}h^{11} +
h_{22}h^{22} + h_{33}h^{33}
\end{equation}
Taking into account that the metric used to raise and lower indices
is the $\gamma$ given in (\ref{gamma}) we observe that only 
the second term of the right hand side of (\ref{finit}) will be
negative, whereas all the others are positive. Now it can easily be
shown, using the explicit solutions (\ref{h00}) and (\ref{h30}),
that the sum $\lambda^2h_{00}h^{00} + \mu^2h_{30}h^{30}$ always admits
a \emph{positive} divergent $\frac{1}{r-r_{hoz}}$ term at the
horizon $r_{hoz} = m + \sqrt{m^2+n^2}$ unless $\lambda=\mu=0$. 
(Remark: The sum may or may not 
have a divergent $\frac{1}{(r-r_{hoz})^2}$ term of whatever sign. 
The parameters $\lambda$ and $\mu$ could
be chosen to cancel it.) Hence the perturbation $h_{ab}$ is
not finite unless we set $h_{00}=h_{30}=0$. From $h_{00}=0$
and equations (B12-B15) of \cite{Young} it follows that $h_{33}=0$ and
$X=h_{11}+h_{22}=0$ which together with $Y=h_{11}-h_{22}=0$ implies
$h_{11}=h_{22}=0$. This finally shows that $h_{ab}=0$ for all $a,b$. 
There are no $SU(2)$ invariant, finite perturbations of Taub-NUT space.

\section{Discussion}
The fact that there are no $SU(2)$-invariant perturbations of
Lorentzian Taub-NUT space is a clear indication that the space is
unstable.  In proving the result one clearly noted how the Lorentzian 
signature of the spacetime enforced possible
perturbations to oscillate near infinity. A straightforward
generalisation would be to consider the cases with non-vanishing
cosmological constant and electric charge, where I expect the same
result to hold. Higher dimensional generalizations also seem possible
with the advanced technical effort of finding the analogues 
of the Wigner-functions. \\
The result is in accord with the result of \cite{Moncrief},
in that it shows pathological behaviour already appears at the
level of a linear stability analysis. \\
It is unlikely that a physically sensible thermodynamics can be 
associated to Lorentzian Taub-NUT spaces and the possible attempts need
to be rethought. In fact the so far unsuccessful efforts to establish
a truely satisfying first law are very likely to be related to 
the instability of the space. Moreover, the linear analysis might help
to gain a better understanding of why standard strategies fail.

\section{Acknowledgements}
I want to thank Gary Gibbons for suggesting the problem and 
useful advice, and Claude Warnick for helpful 
discussions and sharing his notes. 
I am also indebted to the following bodies for 
financial support: EPSRC, Studienstiftung des deutschen
Volkes and the Allen, Meek and Read Scholarship.

\begin{appendix}
\section{NUT-Thermodynamics}
In this appendix I want to show how a Smarr-formula can be derived for
Taub-NUT space via dimensional reduction along the $U(1)$ fibres of
the metric and comment on unsuccessful attempts to deduce a first law
from it. 

\subsection{Komar-Integrals}
Let us compute the conserved quantity associated with the Killing
vector $\frac{\partial}{\partial t}$. The corresponding Killing 
one-form is
\begin{equation}
k=\frac{\Delta(r)}{r^2+n^2} (dt + 2n \cos \theta d \phi )\,.
\end{equation}
Thus up to a sign depending on one's convention for the Hodge
operator $\star$:
\begin{equation}
\star dk= \left(\frac{\Delta(r)}{r^2+n^2}\right)^\prime (r^2 + n^2) \sin \theta d \theta \wedge d \phi
+ \frac{2n}{r^2+n^2} dr \wedge  (dt + 2n \frac{\Delta(r)}{r^2+n^2} \cos \theta d \phi )\,.
\end{equation}
Care is needed when integrating over $\theta$  and $\phi$ at
constant $r$. One can formally perform the integration
\begin{equation}
\int d \star k= \int  \left(\frac{\Delta(r)}{r^2+n^2}\right)^\prime (r^2 + n^2 ) 
\sin \theta d \theta \wedge d \phi = m\frac{r^2 -n^2}{r^2 +n^2 } 
+\frac{2 n^2 r}{r^2 + n^2 } \ ,
\end{equation}
finding that for $n \neq 0$ the integral depends on $r$. Strictly
speaking, one has to take the Misner-strings on the z-axis
(respectively the two different charts) into account when 
performing the integration. Thus one should omit from the 
integral a small region, radius $\epsilon$ around the intersection
of the two concentric spheres with the Misner strings.  
To get a closed 2-cycle, homologically trivial, 
one should add two small tubes surrounding the Misner strings.
The integral over the tubes would make a non-vanishing contribution  
in the limit $\epsilon \downarrow 0$. This limit can be attributed to
the Misner strings. But this means that the difference of the above
integral evaluated at infinity and at the horizon must be the
tube-contribution of the Misner strings. 
At infinity
\begin{equation}
\frac{ 1}{ 8 \pi} \int \star dk \rightarrow m \, ,
\end{equation}
whereas at the Cauchy horizon, $r=r_+ =  m + \sqrt{m^2 +n^2}$, we have
\begin{equation}
\frac{ 1}{ 8 \pi} \int_{r_+} \star dk = \sqrt{m^2+n^2} \ .
\end{equation}
The contribution due to the Misner strings is given by the difference
\begin{equation}\label{nema}
\int _{\rm Misner \,Strings} \star dk = m - \sqrt{m^2+n^2} \ .
\end{equation}
It is negative for $n>0$. The result can be interpreted in that the
spacetime contains negative energy between the horizon at $r_+$ and
infinity. Note that the result is reminiscent of the result of 
\cite{Manko}. These authors associate the negative mass (\ref{nema}) to the 
(in their case intrinsic) line-singularities on the z-axis, whereas we
do not have intrinsic singularities but obviously a negative mass
content in our spacetime.
To make contact with a Smarr-type formula we write the result
(\ref{nema}) in a different way. A calculation of the surface gravity 
of the horizon at $r=r_+$ yields 
\begin{equation}
\kappa = \frac{1}{2} \frac{ r_+-r_- }{ r_+^2+n^2 }
= \frac{\sqrt{m^2 + n^2 }}{ r_+ ^2 + n^2 }\,.
\end{equation}
The area $A$ of the horizon is 
\begin{equation}
A= 4 \pi (r_+^2 + n^2)\, .
\end{equation}
Now obviously
\begin{equation}\label{sma1}
 m = \frac{ \kappa A}{ 4 \pi} + \int _{\rm Misner \,Strings} \star dk\,.
\end{equation}
We can extract more insight in the term which is due to the Misner
strings by using a dimensional reduction.

\subsection{Dimensional Reduction}
The idea of the method of dimensional reduction is to project 
the Einstein equations down to the space $\Sigma$ of orbits of the 
Killing field $\p_t$. Note that $\Sigma$ should not be thought of 
as a submanifold of the spacetime manifold $M$ but -- locally at least --
 as the base space of an ${\Bbb R}$-bundle over $\Sigma$.   
If one focusses on the timelike orbits, the space
of orbits will have a boundary component $\partial \Sigma_+$ 
on which the orbits become null, that is the boundary consists 
of the space of null generators of the Cauchy horizon at $r=r_+$. Of
course, $\Sigma$ will also have a boundary at infinity ($r \rightarrow
\infty$), which we denote $\partial \Sigma_\infty$.

Outside the horizon, the four-dimensional metric may be written  as  
\begin{equation}
ds ^2 = -e^{2U} (dt + \omega) ^2 + e^{-2U} \gamma _{ij} dx ^i dx ^j\,,
\end{equation}
with 
\begin{align}
\omega&=2n \cos \theta d\phi \, , \\
U(r)&=\frac{1}{2} \log \frac{\Delta(r)}{r^2+n^2}  \, , \\
\gamma_{ij}&=diag \left(1, \ e^{2U}(r^2+n^2), \ e^{2U}(r^2+n^2)\sin^2
\theta\right) \, .
\end{align}
Let $\Omega = d\omega$. Then the vacuum Einstein equations imply that 
\begin{equation}
d \star \bigl (e^{4U} \Omega \bigr)=0\,,
\end{equation}
where the Hodge operator $\star$ is taken with respect to the 
metric $\gamma_{ij}$. Thus one may introduce a NUT-potential $\psi$
such that
\begin{equation}
\Omega = e^{-4U} \star d \psi \,.
\end{equation}
 whence
\begin{equation} \label{psi} 
\nabla _i  \bigl ( \gamma ^{ij}   e^{-4U} \nabla _j \psi  )  =0\,.
\end{equation}
The Einstein equations may now be obtained
from the three-dimensional Lagrangian density 
\begin{equation}
\mathcal{L}=\sqrt{\gamma}\left(R + 2 \gamma ^{ij} \partial_i U
\partial _j U +\frac{1}{2} e^{-4U} \gamma^{ij} \partial _i \psi \partial _j \psi\right) \,.
\end{equation}
where $R$ is the Ricci scalar of the metric $\gamma_{ij}$. 
The equation of motion for $U$,
\begin{equation}
\nabla ^2 U + \frac{1}{2} e^{-4U} (\partial \psi)^2=0\, , \label{U} 
\end{equation}
can be solved for the potential $\psi$ with the result
\begin{equation}
\psi = -2n \frac{ m-r}{(r^2+n^2)}=\frac{2n}{ r} +
\mathcal{O}\left(\frac{1}{r^2}\right) \,.
\end{equation}
Now from (\ref{psi}) one may obtain the identity  
\begin{equation}
-8 \pi n= \int_{\partial \Sigma_{\infty} } e^{-4U}  \partial _i \psi d \sigma ^i=
\int_{\partial \Sigma_+ } e^{-4U}  \partial _i \psi d \sigma ^i.
\end{equation}
From (\ref{U}) one may obtain the identity
\begin{equation}
4 \pi m= \int_{\partial \Sigma_{\infty}}  \partial _i Ud \sigma ^i
=  -\frac{1}{2}\int _\Sigma e^{-4U} (\partial \psi)^2 +\int_{\partial \Sigma_+ } 
\partial _i U d \sigma ^i\,.
\een
But
\begin{equation}
\int_{\partial \Sigma_+ } 
\partial _i  U d \sigma ^i =
\frac{1}{2}\frac{\Delta^\prime(r_+)}{\Delta(r_+)}\Delta(r_+)(r^2+n^2)\cdot
4\pi = \kappa  A \,.
\end{equation}
Moreover, multiplication of (\ref{psi}) by $\psi$ and integration by parts
yields
\begin{equation}
-\int_\Sigma  e^{-4U} (\partial \psi) ^2 = \int _{\partial
  \Sigma_+}\psi e^{-4U} \partial _i \psi d \sigma ^i
-\int _{\partial \Sigma_\infty}\psi e^{-4U} \partial _i \psi d \sigma ^i
\, . 
\end{equation}
Note that the boundary term at infinity vanishes.
Putting this all together gives
\begin{equation}
4 \pi m =  \kappa A - 4 \pi n \psi_H \, ,
\end{equation}
and hence 
\begin{equation} \label{smarr}
m = \frac{\kappa A}{4 \pi} - \psi_H n \ ,
\end{equation}
where $\psi _H$ is the value of the NUT-potential
on the horizon. Comparing (\ref{smarr}) with (\ref{sma1}) 
we have found an interpretation for the Misner-term
as a NUT-potential, which is constant on the horizon, multiplied by
the NUT-charge. In fact it is straightforward to generalize
(\ref{smarr}) to the Kerr-Taub-NUT case.

\subsection{Failure of a First Law}
From (\ref{smarr}) one would expect the first law
\begin{equation} \label{fil}
dm \ \ ^{!}= \, ^{!} \ \frac{\kappa dA}{8 \pi} - \psi_H dn
\end{equation}
to hold. However, (\ref{fil}) does not hold unless a special relation 
between $m$ and $n$ is assumed, which the reader might wish to work
out. This condition is -- at least for the general Kerr-NUT-case -- 
different from the regularity condition that follows from the usual 
periodicity-condition that is imposed on the Euclideanized metric. \\
Moreover it can be shown for the Kerr-case \cite{Gustav} 
that there exists no modification of the area that makes a first law
with NUT-potential hold.
\end{appendix}

\end{document}